\documentclass[pre,twocolumn,
superscriptaddress,
showpacs,preprintnumbers]{revtex4}

\usepackage{amssymb}
\usepackage{graphicx}

\begin{document}

\title{Equivalence between the mobility edge of electronic transport on
disorder-less networks and the onset of chaos via intermittency in
deterministic maps}
\author{M. Mart\'{\i}nez-Mares}
\affiliation{Departamento de F\'{\i}sica, Universidad Aut\'onoma
Metropolitana-Iztapalapa, A. P. 55-534, 09340 M\'exico D. F., Mexico}
\author{A. Robledo}
\affiliation{Instituto de F\'{\i}sica, Universidad Nacional Aut\'onoma de M\'exico, 
A. P. 20-364, 01000 M\'exico D. F., Mexico}
\date{\today}

\begin{abstract}
We exhibit a remarkable equivalence between the dynamics of an intermittent
nonlinear map and the electronic transport properties (obtained via the
scattering matrix) of a crystal defined on a double Cayley tree. This strict
analogy reveals in detail the nature of the mobility edge normally studied
near (not at) the metal-insulator transition in electronic systems. We
provide an analytical expression for the conductance as function of system
size that at the transition obeys a $q$-exponential form. This
manifests as power-law decay or few and far between large spike oscillations 
according to different kinds of boundary conditions.
\end{abstract}

\maketitle


Occasionally the detection of a deep running analogy between two apparently
different physical problems allows for the determination of elusive
quantities and understanding of difficult issues. Here we present a
relationship between intermittency and electronic transport. This
development brings together fields of research in nonlinear dynamics and
condensed matter physics. Specifically, the dynamics at the onset of chaos
appears associated to the critical conductance at the mobility edge of
regular self-similar networks \cite{mobility}.

Recently, the dynamics at the transitions to chaos that occurs along the
three known universal routes from regular to irregular behavior (in
low-dimensional nonlinear maps) has been analyzed with a good deal of detail 
\cite{Baldovin2002}-\cite{Robledo2008}. This effort has helped establish
the nature of the statistical-mechanical structure obeyed by the dynamics
associated to nonmixing and nonergodic attractors \cite{Robledo2008}. On the
other hand, there are known connections between nonlinear dynamical systems
and electronic transport properties. For example, there are models for
transport in incommensurate systems, where Schr\"{o}dinger equations with
quasiperiodic potentials \cite{Harper1955} are equivalent to nonlinear maps
with a quasiperiodic route to chaos, and where the divergence of the
localization length translates into the vanishing of the ordinary Lyapunov
coefficient \cite{Ketoja1997}.

At the tangent bifurcation \cite{Baldovin2002}, the focal point of the
intermittency route to chaos, an uncommon but welcome simplicity has led to
analytical results in closed form for the dynamics at vanishing Lyapunov
exponent \cite{Baldovin2002}. Here we make full use of this circumstance
showing that transport in a model network, a double Cayley tree, resolved by
means of the scattering matrix, is given by the properties of a
one-dimensional nonlinear map. The model, in this study, does not
contain disorder; nevertheless it displays a transition between localized
and extended states. The translation of the map dynamical features into
electronic transport terms provides not only the description of the two
different conducting phases but, we believe, offers for the first time a
rigorous account of the conductance at the mobility edge. A type of
localization length in the incipient insulator mirrors the departure from
exponential sensitivity to initial conditions at the transition to chaos.

\begin{figure}[tbp]
\includegraphics[width=6.9cm]{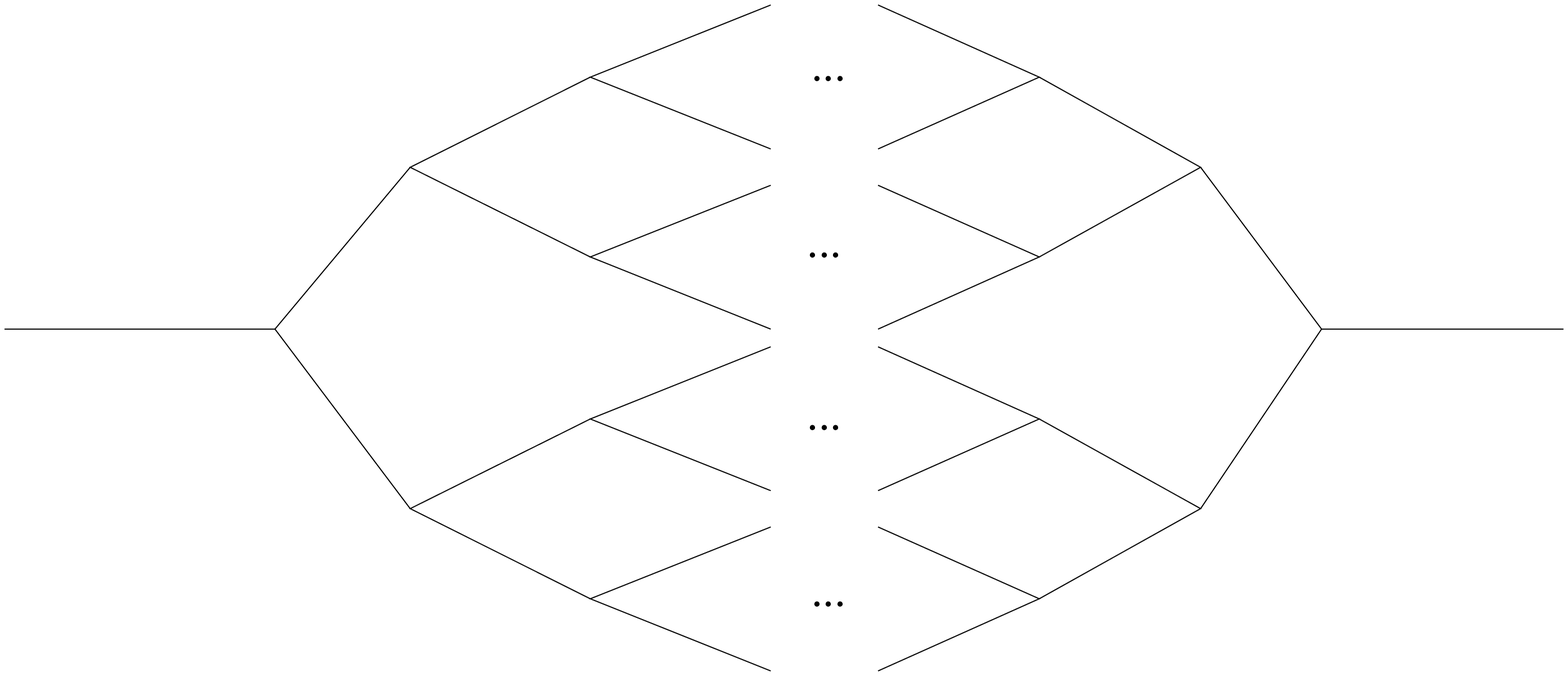}
\caption{A double Cayley tree of conectivity $K=2$ and lattice constan $a$.
Each bond is a perfect one dimensional conductor.}
\label{fig:cayleyd1}
\end{figure}

We recap briefly the usefulness of Cayley tree networks in the study of
electronic transport properties in the presence, and absence, of disorder. A
single Cayley tree spans over a space of infinite dimensionality \cite%
{Straley} and transport on it exhibits a metal-insulator transition as a
function of disorder \cite{abouchacra}. A scattering approach was applied in
Ref. \cite{ShapiroPRL1983} for off-diagonal disorder and shown that the
metal-insulator transition occurs for connectivity $K\geq 2$ ($K+1$ is the
coordination number). A single Cayley tree is a first approximation to an
ordinary regular lattice \cite{abouchacra}, but, as shown below, a double 
Cayley tree (two single Cayley trees joined conformally as in 
Fig.~\ref{fig:cayleyd1}) is a much better approximation (see also 
\cite{Zekri}). In Ref.~\cite{Avishai1992} is shown that the conducting 
band of a disorder-less double Cayley tree contracts and disappears as 
$K$ increases. In Ref.~\cite{Horvat} the dynamic behaviour of a chain of 
scatterers was analyzed in the absence and presence of disorder; while the
localization transition for different types of complex networks, including
the double Cayley tree, was studied in Ref. \cite{Sade} via spectral
statistics. Finally, the double Cayley tree problem is relevant 
for transport in chaotic cavities with broken mirror symmetry 
\cite{GMMB}. Here we show that the electronic structure for $K=2$ can be 
determined by reducing the scattering matrix to a nonlinear map. This 
development facilitates the band description of the conductance as a 
function of energy including the location of the mobility edge.

Here we consider electronic transport in the double Cayley tree (see Fig.~\ref{fig:cayleyd1}). We refer only to the ordered, crystal-like, 
system and reduce its associated scattering matrix to a nonlinear map. 
The number of times the trees are ramified, starting from perfect join, 
is the generation $n$ that quantifies the size of the system. 
Also, for brevity, we will fix the tree connectivity to $K=2$ where one
lead, we call it the incoming lead, is divided into two leads at a given
node. The leads are assumed to be equivalent to one dimensional perfect
wires with length equal to the lattice constant $a$ and are independent of $n
$. Hence, each node is described by the same $3\times 3$ scattering matrix
for which we assume the model \cite{Buettiker1984} 
\begin{equation}
S_{\text{node}}=\left[ 
\begin{array}{ccc}
-(\alpha +\beta ) & \sqrt{\epsilon } & \sqrt{\epsilon } \\ 
\sqrt{\epsilon } & \alpha  & \beta  \\ 
\sqrt{\epsilon } & \beta  & \alpha 
\end{array}%
\right] ,  \label{eq:Snode}
\end{equation}
where $\epsilon $, a real number in the interval 
$0\leq \epsilon \leq \frac{1}{2}$, is the transmission probability (or coupling) 
from the incoming lead
to the others, and viceversa. The reflection amplitude to the incoming lead
is $-(\alpha +\beta )$, with $\alpha =-(1-\sqrt{1-2\epsilon })/2$ and $\beta
=(1+\sqrt{1-2\epsilon })/2$. When incidence is only on one of the other two
leads $\alpha $ is the reflection amplitude to the same lead and $\beta $
the transmission amplitude to the other lead.

The scattering matrix of the system is $2\times 2$ and satisfies a recursive
relation. If we denote by $S_{n}$ the scattering matrix at generation $n$,
the combination rule for scattering matrices allows $S_{n}$ to be written in
terms of the scattering matrix at a previous generation $n-1$, 
\begin{equation}
S_{n}=\frac{-1}{e^{-2ika}\openone-\sqrt{1-2\epsilon }S_{n-1}}
(\sqrt{1-2\epsilon }\,e^{-2ika}\openone-S_{n-1}), 
\label{eq:recursion}
\end{equation}
with $\openone$ the $2\times 2$ identity matrix. The scattering matrix at a
generation $n$ can be obtained iteratively starting from that for the
perfect union in the middle of our double Cayley tree: $S_{0}=\sigma _{x}$,
where $\sigma _{x}$ is a Pauli matrix.

Firstly, it can be seen that $S_{n}$ is a unitary matrix, which is the
condition of flux conservation. Then, time reversal invariance restricts 
$S_{n}$ to be a symmetric matrix. Finally, the additional lattice spatial 
reflection symmetry implies that $S_{n}$ has the form \cite{GMMB} 
\begin{equation}
S_{n}=\left( 
\begin{array}{cc}
r_{n} & t_{n} \\ 
t_{n} & r_{n}
\end{array}
\right) ,
\end{equation}
where $r_{n}$ and $t_{n}$ are the reflection and transmission amplitudes.
With this structure $S_{n}$ is diagonalized by a $\pi /4$-rotation 
\cite{GMMB}; that is 
\begin{equation}
\left( 
\begin{array}{cc}
e^{i\theta _{n}} & 0 \\ 
0 & e^{i\theta _{n}^{\prime }}
\end{array}
\right) =\frac{1}{\sqrt{2}}\left( 
\begin{array}{rc}
1 & 1 \\ 
-1 & 1
\end{array}
\right) S_{n}\frac{1}{\sqrt{2}}\left( 
\begin{array}{cr}
1 & -1 \\ 
1 & 1
\end{array}
\right) .
\label{eq:diagonal}
\end{equation}
Here, $\theta _{n}$ and $\theta _{n}^{\prime }$ are the eigenphases that
satisfy $e^{i\theta _{n}}=r_{n}+t_{n}$ and 
$e^{i\theta _{n}^{\prime}}=r_{n}-t_{n}$. 
In terms of the eigenphases the transmission amplitude is
given by $t_{n}=\frac{1}{2}(e^{i\theta _{n}}-e^{i\theta _{n}^{\prime }})$.
Morever, the dimensionless conductance (i.e. in units of $2e^{2}/h$) depends
on the eigenphases through Landauer's formula as \cite{Landauer,IBM}, 
\begin{equation}
g_{n}=|t_{n}|^{2}.  \label{eq:gn}
\end{equation}
Therefore, the analysis of the eigenphases is of crucial importance as they
determine every transport property.

Central to our discussion is the fact that the recursive 
relation~(\ref{eq:recursion}) can be written in the diagonal form (\ref{eq:diagonal}) and this implies the existence of a one-dimensional 
nonlinear map for the phase $\theta _{n}$. The map 
$\theta _{n+1}=f(\theta _{n})$ can actually be obtained in the following 
closed form 
\begin{eqnarray}
f(\theta _{n}) &=&2ka-\theta _{n}  \nonumber \\
&+&2\arctan \left( \frac{\sin \theta _{n}+\sqrt{1-2\epsilon }\sin 2ka}{\cos
\theta _{n}-\sqrt{1-2\epsilon }\cos 2ka}\right) ,\quad  \label{eq:map-2}
\end{eqnarray}%
where the dependence on $\epsilon $ and $ka$ comes out clearly. In what
follows everything said about $\theta _{n}$ is valid for 
$\theta_{n}^{\prime }$ as well. Perfect union at $n=0$ means 
$\theta _{0}=0$ and $\theta _{0}^{\prime }=\pi $.

\begin{figure}[tbp]
\includegraphics[width=6.9cm]{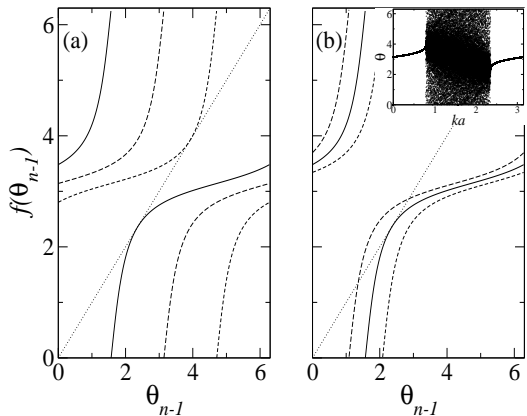}
\caption{$f(\protect\theta _{n})$ of Eq.~(\protect\ref{eq:map-2}) for 
$\protect\epsilon =\frac{1}{4}$: (a) $ka=\protect\pi /4$ (dashed), 
$2\protect\pi /4$ (long dashed), $3\protect\pi /4$ (continuous), and (b) $ka=2.6$
(long dashed), $3\protect\pi /4$ (continuous), $2.1$ (dashed). The dotted
lines correspond to the identity. Inset: Periodic and chaotic attractors.}
\label{fig:map}
\end{figure}

For a given value of $\epsilon $ the map
(\ref{eq:map-2}) is periodic in $ka$ (the parameter related to the energy)
with period $\pi $. In the range $0\leq ka\leq \pi $ the attractor diagram
presents a chaotic region between two windows of period 1 (see inset of 
Fig.~\ref{fig:map}(b)), separated by
bifurcation points at 
\begin{equation}
k_{c}a=\arccos \sqrt{2\epsilon }\quad \mbox{and}\quad k_{c^{\prime }}a=\pi
-\arccos \sqrt{2\epsilon },  \label{eq:critical}
\end{equation}%
such that the chaotic region of the map takes place in the interval 
$k_{c}a<ka<k_{c^{\prime }}a$, while windows of period 1 in $ka\leq k_{c}a$
and $ka\geq k_{c^{\prime }}a$. As we see below Eq. (\ref{eq:critical}) gives
the locations of the mobility edge as a function of the transmission
probability $\epsilon $. Fixed-point solutions $\theta _{\infty }$ for 
$\theta _{n}$, $n\rightarrow \infty $, are 
\begin{equation}
\theta _{\infty }=\left\{ 
\begin{array}{lcc}
\theta _{-} & \mbox{for} & 0\leq ka\leq k_{c}a \\ 
\theta & \mbox{for} & k_{c}a<ka<k_{c^{\prime }}a \\ 
\theta _{+} & \mbox{for} & k_{c^{\prime }}a\leq ka\leq \pi
\end{array}%
\right. ,  \label{eq:infinity}
\end{equation}%
where $\theta =-ka+3\pi /2$, and $\theta _{\pm }$ is given by 
\begin{equation}
\tan \theta _{\pm }=\frac{\sin ka\left( \cos ka\pm \sqrt{\cos
^{2}ka-2\epsilon }\right) }{1-\cos ka\left( \cos ka\pm \sqrt{\cos
^{2}ka-2\epsilon }\right) }.  \label{eq:theta-2}
\end{equation}%
These fixed-point solutions indicate that for large $n$, $\theta _{n}$
reaches the values $\theta _{\pm }$ in the windows of single period, while
in the chaotic region $\theta _{n}$ fluctuates according to an invariant
density with maximum at $\theta $.

At the bifurcation points $k_{c}a$ and $k_{c^{\prime }}a$, the fixed-point
phase $\theta _{\infty }$ takes the critical values 
$\theta _{c}=-\arccos \sqrt{2\epsilon }+3\pi /2$ and 
$\theta _{c'}=\arccos \sqrt{2\epsilon }+\pi /2$, respectively, for 
$\theta _{\infty }$ between $0$ and $2\pi $. 
In Fig.~\ref{fig:map}(a) we plot the map Eq.~(\ref{eq:map-2}) for 
$\epsilon =1/4$ for the three parameter values 
$ka=\pi /4,\,2\pi /4,\,3\pi /4$. 
It is evident that the transitions from chaotic behavior to the windows of
period one are tangent bifurcations as the map is tangent to the identity
line \cite{Baldovin2002} at the critical values $k_{c}a=\pi /4$ and 
$k_{c^{\prime }}a=3\pi /4$, where $\theta _{c}=5\pi /4$ and 
$\theta_{c'}=3\pi /4$. Fig.~\ref{fig:map}(b) shows three cases: secant 
($ka=2.6$) and tangent ($ka=3\pi /4$) period one solutions, and a bottleneck 
($ka=2.1$) that gives rise to intermitency as a precursor to periodic behavior.

Information about transport can be obtained from the sensitivity to initial
conditions that characterizes the dynamics of the nonlinear map. For \emph{%
finite} $n$ it is given by 
\begin{equation}
\Xi _{n}\equiv \left\vert \frac{d\theta _{n}}{d\theta _{0}}\right\vert
\equiv \exp (\Lambda _{1}(n)\ n),  \label{eq:sensitivity}
\end{equation}
where $\theta _{0}$ is an initial condition and the exponential law after
the 2nd identity defines the \emph{finite} $n$ Lyapunov exponent 
$\Lambda_{1}(n)$. For $n$ large $\Lambda _{1}(n)$ becomes the Lyapunov exponent 
$\lambda _{1}$, a number independent of $\theta _{0}$ that according to its
sign characterizes periodic and chaotic attractors. At the tangent
bifurcation $\lambda_{1}=0$ and the sensitivity adopts instead a 
$q$-exponential form (see below) \cite{Baldovin2002,qexponential}. From Eqs. (\ref{eq:gn}), (\ref{eq:sensitivity}) and Eq. (\ref{eq:map-2}) 
(and $t_{n}=\frac{1}{2}(e^{i\theta _{n}}-e^{i\theta _{n}^{\prime }})$) 
we obtain the recursion formula
\begin{eqnarray}
g_{n} & = & g_{n-1}\exp (\Lambda _{1}(n)\ n)\exp (\Lambda _{1}^{\prime }(n)\ n),
\label{eq:gn-recursion} \\
\Lambda _{1}(n) & = & \ln \frac{\epsilon }{1-\epsilon -\sqrt{1-2\epsilon }\cos
(\theta _{n}+2ka)},  \label{eq:lyapunov1}
\end{eqnarray}
where $\Lambda _{1}^{\prime }(n)$ is given by Eq. (\ref{eq:lyapunov1})
with $\theta _{n}$ replaced by $\theta _{n}^{\prime }$. We note that 
$\Lambda _{1}$ and $\Lambda _{1}^{\prime }$, and hence $g_{n}$, do not depend
on the initial conditions $\theta _{0}$ and $\theta_{0}^{\prime }$. The
Lyapunov exponent $\lambda_{1}$ is given by Eq. (\ref{eq:lyapunov1}) with 
$\theta _{n}$ replaced by $\theta_{\infty }$.

\begin{figure}[tbp]
\includegraphics[width=6.8cm]{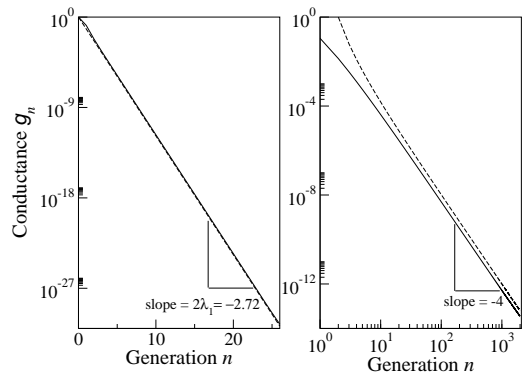}
\caption{Conductance as a function of generation 
for $\protect\epsilon =\frac{1}{4}$: $ka=0.5$ (left panel) and $ka=\protect\pi /4$ 
(right panel). Continuous lines represent $g_{n}$ obtained directly from the map through
Eq.~(\protect\ref{eq:gn}) while dotted lines correspond to 
$g_{n}=\exp (2\protect\lambda_{1}n)$ and $g_{n}=\exp _{3/2}(2\protect\lambda_{3/2}n)$
for left panel and right panel, respectively. The two curves in (b) differ because of a proportionality factor, See Eq. (14).}
\label{fig:gn}
\end{figure}

The dynamical properties of the map (\ref{eq:map-2}) translate into the
following network properties: i) In relation to the attractors of period one
that take place along $0\leq ka<k_{c}a$ and $k_{c^{\prime }}a<ka\leq \pi $,
we corroborate from Eqs. (\ref{eq:infinity}) and (\ref{eq:lyapunov1}) that 
$\lambda_{1}$ is negative and therefore the conductance $g_{n}$ decays
exponentially with system size $n$, $g_{n}=\exp (2\lambda _{1}n)$, implying
localization, the localization length being $\zeta _{1}=a/|\lambda _{1}|$.
In the left panel of Fig. \ref{fig:gn} we see a clear exponential decay of 
$g_{n}$ as a function of $n$ at $ka=0.5$, where we compare $g_{n}$ computed 
directly from Eq.~(\ref{eq:gn}) with that obtained from $\lambda _{1}$. We 
notice that the conductance for the localized states of an ordered system displays 
the same behavior as that in the insulating regime of a disordered wire in
quasi-one-dimensional configuration \cite{Beenakker1997}. ii)
With respect to the chaotic attractors that occur in the interval $%
k_{c}a<ka<k_{c^{\prime }}a$ we observe that $\lambda _{1}$ becomes positive
and the recursion relation Eq. (\ref{eq:gn-recursion}) does not let $g_{n}$
decay but makes it oscillate with $n$ (not shown here) indicating that
conduction takes place. In our model $g_{n}$ does not scale with system size
as in the metallic regime of quasi-one-dimensional disordered wire where
Ohm's law is satisfied \cite{Beenakker1997}. In the parameter region where
the map is incipiently chaotic, say $ka\gtrsim k_{c}a$, the network grows
with $n$ with an insulator character, but interrupted for other intermediate
values of $n$ by conducting crystals. In the map dynamics these are the
laminar episodes separated by chaotic bursts in intermittent trajectories.

The most distinct outcome of our treatment is the description obtained of
the mobility edge from the dynamics at the critical attractors located at $%
ka=k_{c^{\prime }}a$ and $ka=k_{c}a$. There $\lambda _{1}=0$ and according
to Eq. (\ref{eq:gn-recursion}) not much can be said about the size
dependence of the conductance when $n\gg 1$. However we can use to our
advantage the known properties of the anomalous dynamics occurring at these
attractors once they are identified as tangent bifurcations [see Fig.~\ref%
{fig:map}(b) for $ka=3\pi /4$]. At a tangent bifurcation of general
nonlinearity $z>1$ the sensitivity obeys a $q$-exponential law for large $n$ 
\cite{Baldovin2002}, 
\begin{equation}
\xi _{n}=\Xi _{n\gg 1}=\exp _{q}(\lambda _{q}n)\equiv \lbrack 1-(q-1)\lambda
_{q}n]^{\mp \frac{1}{q-1}},  \label{eq:q-sensitivity}
\end{equation}%
where $\lambda _{q}$ is a $q$-generalized Lyapunov coefficient given by $%
\lambda _{q}=\mp zu$, $q=2-1/z$, where $u$ is the leading term of the
expansion up to order $z$ of $\theta _{n}$ close to $\theta _{c}$ (or $%
\theta _{c^{\prime }}$); i.e. $\theta _{n}-\theta _{c}=(\theta _{n-1}-\theta
_{c})+u\,|\theta _{n-1}-\theta _{c}|^{z}+\cdots $. The minus and plus signs
in Eq. (\ref{eq:q-sensitivity}) and in $\lambda _{q}$ correspond to
trajectories at the left and right, respectively, of the point of tangency 
$\theta _{c}$. Eq. (\ref{eq:q-sensitivity}) implies power-law decay of 
$\xi_{n}$ with $n$ when $\theta _{n}-\theta _{c}<0$ and faster than exponential
growth when $\theta _{n}-\theta _{c}>0$. (We recall that any choice of $S_{0}
$ other that the Pauli matrix translates into another initial condition for
the map). By making the expansion around $\theta _{c}$ (or $\theta_{c'}$) 
for our map (\ref{eq:map-2}) we find (as evidently anticipated) $z=2$ implying 
$q=3/2$, $u=\sqrt{(1-2\epsilon )/2\epsilon }$, and the $q$-generalized Lyapunov 
exponent is $\lambda_{3/2}=-2\sqrt{(1-2\epsilon)/2\epsilon}$. Following the same 
steps that lead to Eq. (\ref{eq:gn-recursion}), the recursion relation for 
$g_{n}$ at each bifurcation point takes the form  
$g_{n}=g_{n-1}\exp _{3/2}(\Lambda_{3/2}(n)n)\exp _{3/2}(\Lambda _{3/2}^{\prime }(n)n)$, so that when 
$\theta_{n}-\theta _{c}<0$ (see Eq. (\ref{eq:q-sensitivity})), 
$g_{n}=\exp _{3/2}(2\lambda _{3/2}n)$, or 
\begin{equation}
g_{n} \propto \left( 1-\frac{1}{2}\lambda _{3/2}n\right) ^{-4}.
\label{eq:transition}
\end{equation}
In the right panel of Fig.~\ref{fig:gn} we compare the results
from Eq.~(\ref{eq:gn}) (continuous line) and Eq.~(\ref{eq:transition})
(dashed line). It is clear that $g_{n}$ decays as a power law (with quartic
exponent) rather than the exponential in the insulating phase. We enphasize
that a localization length given by $\zeta _{3/2}=a/\lambda _{3/2}$ can
still be defined at the mobility edge. To our knowledge this property has
not been reported before. When $\theta _{n}-\theta _{c}>0$, 
$\lambda _{3/2}=2\sqrt{(1-2\epsilon)/2\epsilon }>0$ and the recursion 
relation for $g_{n}$ describes, as the result of the diverging  
duration of the laminar episodes of intermittency, large $n$ intervals  
of vanishing $g_n$ between increasingly large spike oscillations.

In summary, we can draw significant conclusions about electronic transport
from our study. These arise naturally when considering the dynamical
properties of the equivalent nonlinear map near or at the intermittency
transition to chaos. Since iteration time in the map translates into the
generation $n$ of the network, time evolution means growth of system size,
reaching the thermodynamic limit (and true self similarity) when 
$n\rightarrow \infty $. In that limit, windows of period one separated by a
chaotic band correspond, respectively, to localized and extended electronic
states. Further, in the referred parameter 
($ka$, $\epsilon$) regions the conductance $g_{n}$ of the model crystal shows
either an exponential decay with system size, with localization length given
by $\zeta_{1}$ (as in the case of a quasi-one-dimensional disordered wire 
\cite{Beenakker1997}), or an oscillating property signalling conducting
states. The pair of tangent bifurcation points of the map correspond to the
band or mobility edges that separate conductor from insulator behavior. At
these bifurcations the sensitivity to initial conditions $\xi _{n}$
exhibits either power-law decay (when $\theta_{0}<\theta _{c'}$ or 
$\theta_{0}>\theta_{c}$) or faster than exponential increase 
($\theta_{c'}<\theta_{0}<\theta_{c}$) and consequently
the conductance inherits comparable decay or variability with system size 
$n$. Notably, as we have seen we can still define a localization length, the 
$q$-generalized localization length $\zeta _{q}$ with a fixed value of $q=3/2$.
This expression is universal, i.e. it is satisfied by all maps that in the
neighborhood of the point of tangency have quadratic term, i.e. $z=2$ 
\cite{Baldovin2002}.
This quantity can be obtained directly by evaluation of $g_n$ 
(in Eq. (\ref{eq:gn})) when $n\rightarrow\infty$. At the mobility edge
$\zeta_1^{-1}=-\lim_{n\rightarrow\infty} n^{-1}\ln g_n$ vanishes because 
$\ln g_n$ no longer decreases linearly with $n$, as it is the case in the 
insulating phase. However, use of 
$\zeta_q^{-1}=-\lim_{n\rightarrow\infty}n^{-1}\ln_q g_n$ leads
to a finite number for one particular value of $q$, $q=3/2$, when the
degree of deformation $q$ in the $q$-logarithm restores linear
behavior. Otherwise $\zeta_q$ vanishes or diverges.

In spite of the unusual features of the double Cayley tree transport model, the 
complete set of exact solutions derived from it provides a comprehensive picture 
about non-exponential behavior of central quantities like the conductance at the 
transition between the insulator and conductor regimes. 

We are indebted to P. A. Mello for pointing out and introducing us to the
model and techniques to study the mobility edge presented here. AR
recognizes support by DGAPA-UNAM and CONACYT (Mexican agencies).



\begin{thebibliography}{99}
\bibitem{mobility} We chose to call here mobility edge, usually employed in
reference to disordered systems, the transition between localized and
extended electronic states in regular self-similar networks.

\bibitem{Baldovin2002} F. Baldovin, A. Robledo, Europhys. Lett. \textbf{60},
518 (2002).

\bibitem{Mayoral2005} E. Mayoral, A. Robledo, Phys. Rev. E \textbf{72},
026209 (2005).

\bibitem{Robledo2008} A. Robledo, L. G. Moyano, Phys. Rev. E \textbf{77},
036213 (2008).

\bibitem{Harper1955} P. G. Harper, Proc. Phys. Soc. London, Sect. A \textbf{68}, 874 (1955).

\bibitem{Ketoja1997} J. A. Ketoja, I. I. Satija, Physica D \textbf{109}, 70
(1997).

\bibitem{Straley} J. P. Straley, J. Phys. C \textbf{10}, 3009 (1977).

\bibitem{abouchacra} R. Abou-Chacra, P. W. Anderson, D. J. Thouless, J.
Phys. C \textbf{6}, 1734 (1973).

\bibitem{ShapiroPRL1983} B. Shapiro, Phys. Rev. Lett. \textbf{50}, 747
(1983).

\bibitem{Zekri} N. Zekri, A. Brezini, Phys. Stat. Solidi B \textbf{133}, 463 (1986).

\bibitem{Avishai1992} Y. Avishai, J. M. Luck, Phys. Rev. B \textbf{45}, 1074
(1992).

\bibitem{Horvat} M. Horvat, T. Prosen, J. Phys. A: Math Theor. \textbf{40},
11593 (2007).

\bibitem{Sade} M. Sade, T. Kalisky, S. Havlin, R. Berkovits, Phys. Rev. E 
\textbf{72}, 066123 (2005).

\bibitem{GMMB} M. Mart\'{\i}nez, P. A. Mello, Phys. Rev. E
 \textbf{63}, 016205 (2000).

\bibitem{Buettiker1984} M. B\"{u}ttiker, Y. Imry, M. Ya. Azbel, Phys. Rev. A 
\textbf{30}, 1982 (1984).

\bibitem{Landauer} R. Landauer, J. Phys.: Condens. Matter \textbf{1}, 8099
(1989).

\bibitem{IBM} M. B\"{u}ttiker, IBM J. Res. Dev. \textbf{32}, 317 (1988).

\bibitem{qexponential} The $q$-exponential and its inverse the 
$q$-logarithm are defined, respectively, as 
$\exp_q(x)\equiv[1+(1-q)x]^{1/(1-q)}$ for $x, q\in\mathbb{R}$, and 
$\ln_q(y)\equiv(y^{1-q}-1)/(1-q)$ for $y\in\mathbb{R}^{+} $. The ordinary 
exponential and logarithm are recovered when $q=1$.

\bibitem{Beenakker1997} C. W. J. Beenakker, Rev. Mod. Phys. \textbf{69}, 731
(1997).

\end{thebibliography}
\end{document}